\documentclass[twocolumn,showpacs,amsmath,amssymb,prl,superscriptaddress,floatfix,groupedaddress]{revtex4-1}
\usepackage{graphicx}
\usepackage{bm}
\usepackage{amsmath,amsfonts}

\begin{document}

\title{Kinetic constraints, hierarchical relaxation and onset of glassiness in strongly interacting and dissipative Rydberg gases}

\author{Igor Lesanovsky}
\author{Juan P. Garrahan}
\affiliation{School of Physics and Astronomy, University of
Nottingham, Nottingham, NG7 2RD, UK}


\date{\today}

\begin{abstract}
We show that the dynamics of a laser driven Rydberg gas in the limit of strong dephasing is described by a master equation with manifest kinetic constraints. The equilibrium state of the system is uncorrelated but the constraints in the dynamics lead to spatially correlated collective relaxation reminiscent of glasses. We study and quantify the evolution towards equilibrium in one and two dimensions, and analyze how the degree of glassiness and the relaxation time are controlled by the interaction strength between Rydberg atoms. We also find that spontaneous decay of Rydberg excitations leads to an interruption of glassy relaxation that takes the system to a highly correlated non-equilibrium stationary state. The results presented here, which are in principle also applicable other systems such as polar molecules and atoms with large magnetic dipole moments, show that the collective behavior of cold atomic and molecular ensembles can be similar to that found in soft condensed-matter systems.
\end{abstract}

\maketitle

The probing and understanding of matter in and out of equilibrium is to date one of the biggest challenges in physics \cite{Hinrichsen2000,*Binder2011,*Peliti2011,*Seifert2012,[{See for example, }]Diehl2010, *Polkovnikov2011,*yukalov2011,*Calabrese2011,*Trotzky2012,*Gambassi2012,*Caux2013}. One currently very successful platform for the exploration of many-body quantum systems are gases of ultra cold atoms which are nowadays routinely prepared and studied in the laboratory \cite{Bloch2008}. While initially these experiments focused on atoms in their electronic ground states there is currently a shift of emphasis towards atoms in highly excited states---so-called Rydberg atoms \cite{Gallagher84,*Saffman10}. These atoms are long lived and interact strongly over long distances thus permitting the exploration of a wide range of many-body phenomena in strongly interacting systems. This has led to a growing body of experimental \cite{Low09,Viteau11,Schwarzkopf2011,Schauss12,Lochead2013,Hofmann2013} and theoretical work \cite{Ates07,Sun08,Weimer08,Weimer2010,Pohl2010,Lesanovsky2011,Ji2011,Dudin2012,Garttner2012,Ates12-2} centered around the exploration of the statics and dynamics of interacting Rydberg gases. With regards to the latter in particular the interplay between dissipation, interaction and coherent laser excitation was recently shown to lead to intriguing dynamical phase transitions and to the formation of ordered stationary states \cite{Ates06,Lee11,*Lee12-1,*Lee12-2,Honing13,*Petrosyan13-1,*Petrosyan13-2,*Petrosyan13-3}.

In this work we focus on a new dynamical aspect of a strongly interacting Rydberg lattice gas. We show that in the limit of strong dissipation this system realizes a glass \cite{[{For a recent review see: }]Biroli2013} in these sense that its stationary state is trivial, i.e.\ completely mixed, but the non-equilibrium relaxation towards it is strongly nontrivial due to kinetic constraints \cite{Ritort2003}.  We show that these constraints---which are typically at the heart of rather idealized glass models \cite{Ritort2003,Garrahan2002,*Garrahan2003}---appear naturally in the effective evolution equation of the Rydberg system. They manifest themselves in characteristic dynamical features such as a dramatic increase of the relaxation time scale of certain arrangements of particles and the occurrence of dynamical heterogeneity \cite{Biroli2013,[{For reviews see, for example, }]Ediger2000,*Berthier2011,*Chandler2010}. We discuss these effects in detail in one and two dimensional settings which are relevant to current experimental efforts. Our work demonstrates that Rydberg gases are not only interesting for the exploration of coherent effects in strongly interacting quantum systems but also because their collective properties can be similar to those of soft matter systems.

\begin{figure}
\includegraphics[width=\columnwidth]{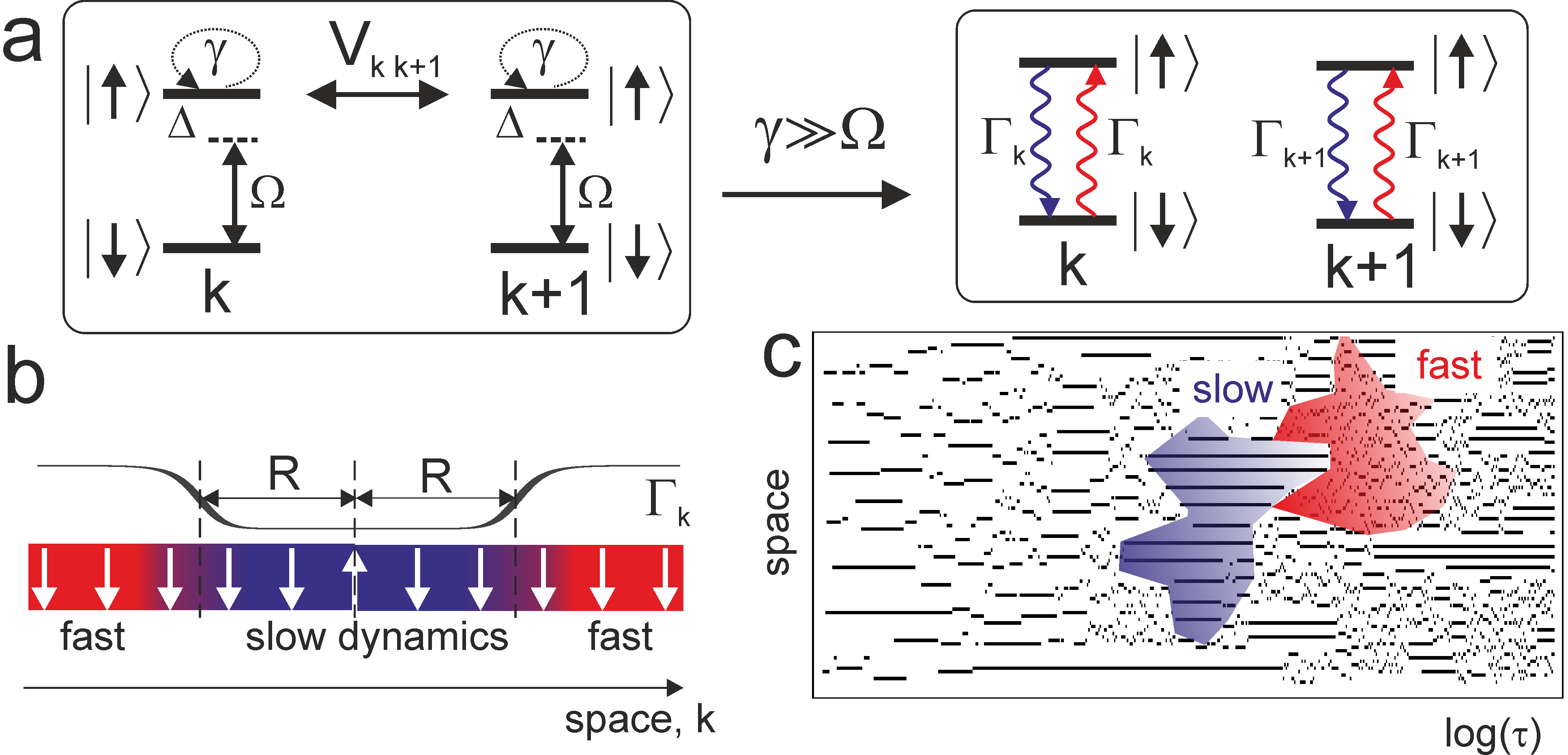}
\caption{(a) Interacting two-level atoms. The ground state $\left|\downarrow\right>$ is coupled to the Rydberg state $\left|\uparrow\right>$ by a laser with Rabi frequency $\Omega$ and detuning $\Delta$. Atoms in Rydberg states interact with the potential $V_{km}$. The ground state dephases with respect to the Rydberg state at a rate $\gamma$. For large $\gamma$ the effective atomic dynamics is described by incoherent state changes at an operator valued rate $\Gamma_k$. (b) The rate $\Gamma_k$ is determined by the number of Rydberg atoms in the vicinity of the $k$-th spin. The dynamics of atoms positioned within a certain distance to an up-spin---characterized by the interaction parameter $R$---is dynamically constrained and therefore slow. More distant atoms evolve rapidly. (c) Trajectory of a one-dimensional system with $N=100$ atoms and $R=8$. The kinetic constraint leads to the emergence of dynamical heterogeneity which manifests in space-time bubbles of fast or slow dynamics.}
\label{fig:system}
\end{figure}

We consider $N$ atoms that are located at the sites of a regular linear or square lattice with spacing $a$ \cite{Schauss12}. For modeling the internal degrees of freedom of the atoms we employ the common spin $1/2$-description (see Fig. \ref{fig:system}a), where the Rydberg state and the ground state are denoted by $\left|\uparrow\right>$ and $\left|\downarrow\right>$, respectively. The two states are coupled by a laser with Rabi frequency $\Omega$ and detuning $\Delta$ with respect to the $\left|\downarrow\right>\rightarrow \left|\uparrow\right>$ transition. Atoms in Rydberg states located at sites $k$ and $m$ (with position vectors $\mathbf{r}_k$ and $\mathbf{r}_m$, respectively) interact strongly with a power law potential $V_{km}=C_\alpha/|\mathbf{r}_k-\mathbf{r}_m|^\alpha$. Depending on the choice of Rydberg states we have either $\alpha=3$ (dipole-dipole interaction) or $\alpha=6$ (van der Waals interaction). Here we focus mostly on the latter case.  In addition, we consider decoherence due to laser phase noise which leads to the dephasing of the $\left|\uparrow\right>$ with respect to the state $\left|\downarrow\right>$ at rate $\gamma$.

The dynamics of the atoms in the lattice is governed by the master equation $\partial_t\rho=\mathcal{L}_0\rho+\mathcal{L}_1\rho$ with $\mathcal{L}_0\rho = -i\left[H_0,\rho\right] +\gamma\sum^N_{k=1} \left( n_k\rho n_k-\frac{1}{2}\left\{n_k,\rho\right\} \right)$ and $\mathcal{L}_1\rho = -i\Omega\sum^N_{k=1}\left[ \sigma_x^k,\rho\right]$. Here $n_k=(1-\sigma_z^k)/2$ and $H_0=\sum^N_{k=1}\Delta n_k+\sum_{km}\frac{V_{km}}{2}n_k n_m$ with $\sigma^k_{x,y,z}$ being the Pauli matrices.

We are interested in the regime in which $\mathcal{L}_0$ is dominant which is the case when the dephasing rate is large, $\gamma\gg |\Omega|$. We therefore ``integrate out'' the fast dynamics governed by the operator $\mathcal{L}_0$ containing the dissipation stemming from dephasing. To this end we define the projection operator $\mathcal{P}=\lim_{t\to\infty} e^{\mathcal{L}_0\,t}$ which projects on the stationary subspace of the fast (dephasing) dynamics in which then the slow effective dynamics takes place. The respective density matrix within this subspace is obtained by the projection $\mu\equiv\mathcal{P}\rho$. This projected density matrix $\mu$ is diagonal in the $\left|\uparrow\right>,\left|\downarrow\right>$-basis. Up to second order in the perturbation $\mathcal{L}_1$ it evolves under the effective master equation $ \partial_t\mu=\int_0^\infty dt\,\mathcal{P} \mathcal{L}_1 e^{\mathcal{L}_0\,t} \mathcal{L}_1 \mu\equiv\mathcal{L}_\mathrm{eff}\mu$ with the explicit form (see Supplemental material)
\begin{equation}
  \partial_\tau\mu=\sum_k\Gamma_k\left(\sigma_x^k\mu\sigma_x^k-\mu\right),\label{eq:L_eff}
\end{equation}
where for convenience we have introduced a rescaled time $\tau=(4\Omega^2/\gamma)\times t$. A crucial feature of (\ref{eq:L_eff}) is that the rate $\Gamma_k$ is operator valued
\begin{eqnarray}
  \Gamma^{-1}_k=1+\left[\delta+R^{\alpha}\sum_{k\neq m} \frac{n_m}{|\hat{\textbf{r}}_k-\hat{\textbf{r}}_m|^\alpha}\right]^2,\label{eq:rate}
\end{eqnarray}
where $\delta=\Delta/\gamma$ is the scaled detuning, $R= a^{-1}\,[2\,C_\alpha/\gamma]^{1/\alpha}$ is a parameter controlling the interaction strength and $\hat{\textbf{r}}_m=\textbf{r}_m/a$.
The rates $\Gamma_k$ encode the interaction between atoms through an effective scaled detuning $\delta+R^{\alpha}\sum_{k\neq m} \frac{n_m}{|\hat{\textbf{r}}_k-\hat{\textbf{r}}_m|^\alpha}$, which is the single atom detuning plus the (scaled) energy shift caused by other Rydberg atoms. This --- so-called kinetic constraint \cite{Ritort2003,Garrahan2002}--- means that the dynamics in the vicinity of excited atoms is governed by small rates and is therefore slow (see Fig. \ref{fig:system}b). Such operator-valued rates were introduced \textit{ad hoc} in other works (e.g.  Refs. \cite{Ates06,Petrosyan13-1,Petrosyan13-2,Petrosyan13-3}) in order to sample the stationary state of a driven Rydberg gas via classical Monte Carlo simulations.

The stationary density matrix $\mu_\mathrm{s}$ of the effective evolution equation (\ref{eq:L_eff}) is determined by $\mathcal{L}_\mathrm{eff}\mu_\mathrm{s}=0$ which is solved by the completely mixed state $\mu_\mathrm{s} \equiv 2^{-N}\bigotimes_k {\mathbb I}$: The rates $\Gamma_k$ solely enter as prefactors of the local terms $\sigma_x^k\mu\sigma_x^k-\mu$ which individually equate to zero when $\mu=\mu_\mathrm{s}$. Hence, the stationary state of this interacting many-body system, as described by Eq.\ (\ref{eq:L_eff}), is the same as for non-interacting atoms.

A first insight into the relaxation dynamics can be gained by a simple mean field analysis. In the following we set the detuning $\Delta=0$ and use Eq.\ (\ref{eq:L_eff}) to derive the equation of motion for the density of Rydberg atoms $p_j(\tau)=\left<n_j\right>(\tau)$ which is $\partial_t p_j =\left<\Gamma_j(1-2n_j)\right>$. This equation is not closed, but we can approximate it by replacing expectation values of products of operators by products of expectation values of operators. Assuming furthermore a homogeneous system, i.e. $p_k(\tau)\rightarrow p(\tau)$, we find the mean field equation for the Rydberg density
\begin{eqnarray}
  \partial_\tau p(\tau)=\frac{1-2p(\tau)}{1+[F_\alpha\,R^{\alpha} p(\tau)]^2} , \label{eq:mean-field}
\end{eqnarray}
where the value of the sum $F_\alpha \equiv \sum_k |\hat{\mathbf{r}}_k|^{-\alpha}$ depends on the dimensionality of the system. In one dimension $F_\alpha$ is twice the Riemann zeta function. The stationary value of the mean field density is $p_{\rm s} \equiv \lim_{\tau\rightarrow\infty} p(\tau)=1/2$ which is compatible with a fully mixed stationary state. Integrating the mean field equation with the initial condition $p(0)=0$ leads to the following implicit equation for $p(\tau)$: $\tau=-(1/8)[4+F^2_\alpha R^{2\alpha}]\log(1-2p(\tau))-(1/4)F^2_\alpha R^{2\alpha}(p(\tau)+1)\,p(\tau)$. We can now identify three different regimes: (i) For short times, where $p(\tau)\ll 1/2$, the logarithm and the term quadratic in $p(\tau)$ are negligible and we find $p(\tau)=e^{-\tau}\sinh(\tau)$ and hence a regime where the number of excitations increases exponentially and independently of $R$ at the (fast) rate $\sim 1$. This is due to the initial creation of distant independent Rydberg excitations. (ii) For long times the logarithm dominates. Here the density is approximately $1/2$ and we obtain $p(\tau)\approx [1-\exp(-8\,\tau/(4+ F^2_\alpha R^{2\alpha}))]/2$. The relaxation is again exponential but strongly depends on the interaction parameter $R$. For $R>1$ we expect a dramatic slow down of the dynamics with the equilibration time scaling as $\tau_\mathrm{eq}\sim R^{2\alpha}F^2_\alpha$. (iii) At intermediate times and for large interaction parameter $R>1$ there is a regime in which the density is small $p(\tau)\ll1/2$ but the product $R^{\alpha} p(\tau)$ is much larger than one. Here the solution of the mean field equation is $p(\tau)\sim [3/(R^{2\alpha} F^2_\alpha)]^{1/3}\, \tau^{1/3}$ with an increase of the density of Rydberg atoms which is algebraic in time. In summary, from the mean field treatment and for sufficiently large $R$ we expect to observe a fast exponential increase of the Rydberg density, followed by an algebraic growth and a final very slow exponential approach to the equilibrium density.

In the mean field description all atoms are considered to be equivalent. However, it is known that systems with kinetic constraints---even though their stationary state might be trivial---exhibit a strongly correlated dynamics. This is in particular the case for constrained spin systems used to model glasses \cite{Ritort2003,Garrahan2002,*Garrahan2003,Chandler2010}. In the following we perform a numerical analysis of the relaxation behavior of the Rydberg gas in which Rydberg atoms interact via a van der Waals potential, i.e. $\alpha=6$. This indeed reveals a rich interplay between spatial and temporal fluctuations which is not captured by mean field and which is strongly reminiscent of glassy systems.

A first example of such non-trivial spatio-temporal dynamics is depicted in Fig.\ \ref{fig:system}c where we show a trajectory of a system of $N=100$ atoms and with interaction parameter $R=8$. Starting from the state with zero Rydberg atoms the short time dynamics is governed by the creation of spatially separated Rydberg atoms with a large ``excluded volume'' of size $\sim R\,a$ surrounding each atom. Moreover, we observe a clearly hierarchical relaxation in the sense that this "excluded volume" decreases and the excitation density increases monotonously with time. A second feature displayed by the trajectory---which is typical for glassy systems \cite{Garrahan2002,*Garrahan2003}---is dynamical heterogeneity \cite{Ediger2000}. That means that the relaxation towards (and also in) the stationary state is characterized by large spatial fluctuations in relaxation time scales.  This gives rise to the formation of space-time ``bubbles'' \cite{Garrahan2002,*Garrahan2003}, such as the regions low activity (striped) or high activity (irregular) in the trajectory of Fig.\ \ref{fig:system}c.

\begin{figure}
\includegraphics[width=0.9\columnwidth]{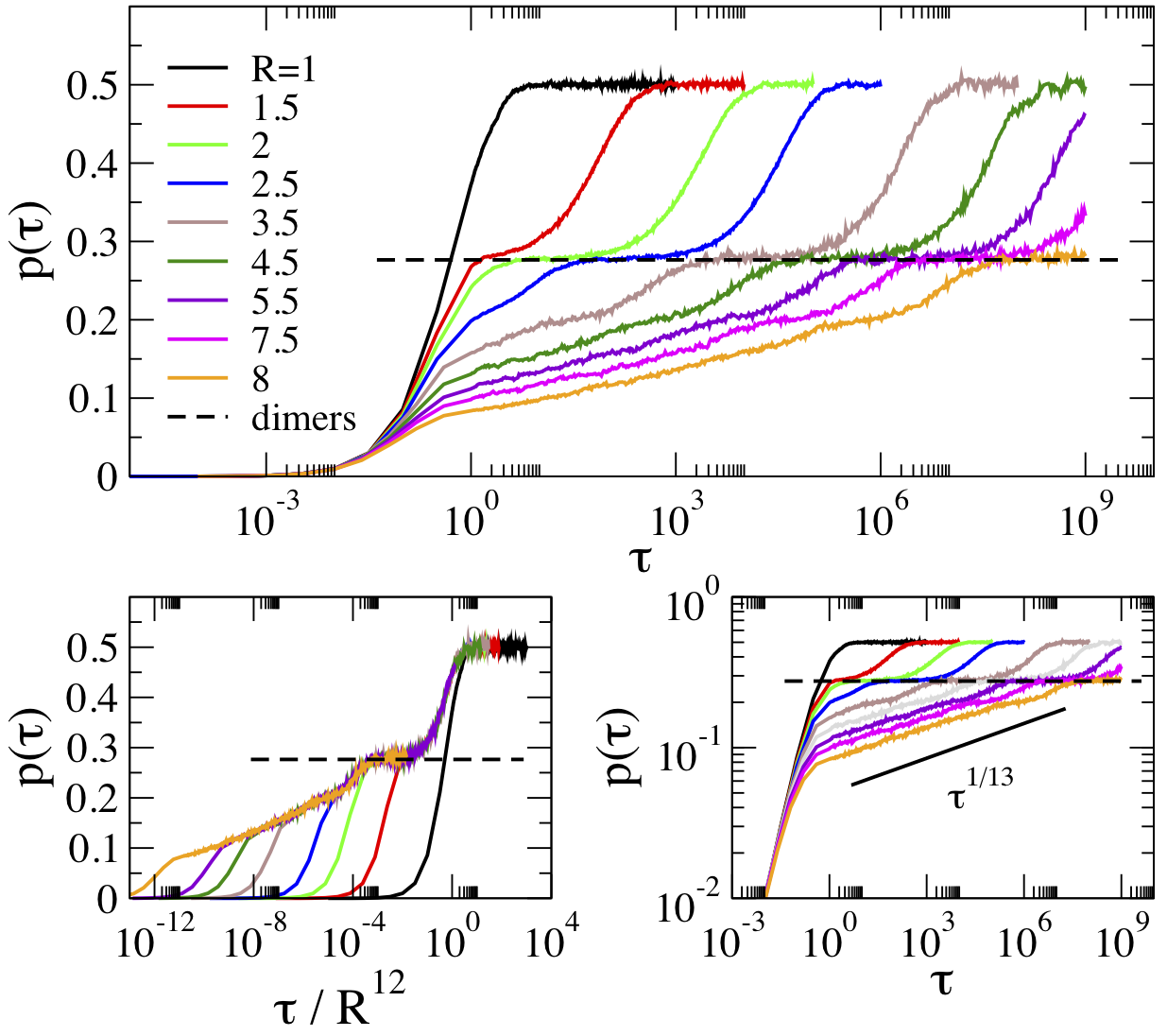}
\caption{Relaxation of the density of Rydberg atoms $p(\tau)= \sum_k p_k(\tau)/N$ in a one-dimensional system with $N=10^5$ atoms, $\alpha=6$ and various values of the interaction parameter $R$. For small times the increase in density is independent of $R$ (top panel) and exponential in $\tau$. For long times the stationary state is approached exponentially at a rate $\propto R^{-12}$ as shown in the bottom left panel. At intermediate times one observes an algebraic growth of $p_k(\tau)$ with an exponent close to $1/13$ (bottom right panel).}
\label{fig:1d_density}
\end{figure}

Let us now study in more detail the relaxation of the mean density $p(\tau)= \sum_k p_k(\tau)/N$ from an initial state with $p_k(0)=0$ as a function of the interaction parameter $R$. The corresponding data is shown in Fig.\ \ref{fig:1d_density}. For the weakly interacting system, i.e. $R=1$, we observe a fast exponential decay to the equilibrium density. As $R$ increases the overall relaxation time increases $\propto R^{12}$ confirming the scaling found from the mean field treatment. The first effect beyond mean field is the emergence of a plateau at a density $p_\mathrm{plat}\approx 0.27$ just before the system relaxes exponentially to the stationary state. On the timescale at which the plateau is present the simultaneous excitation of neighboring spins is strongly suppressed. In this transient state the system is in a mixed state of all configurations in which no nearest neighboring atoms are excited. The density of such state is that of hard dimers at fugacity $1$ which evaluates to $p_\mathrm{dim}=(1-1/\sqrt{5})/2\approx 0.276$ \cite{Lesanovsky2011}. The second effect which goes beyond the mean field treatment concerns the regime of intermediate times, where $R^{\alpha}p(\tau)\gg 1$ but $p(\tau)\ll 1/2$. The simple mean field analysis above suggests an algebraic increase of the Rydberg density with an exponent $1/3$. The numerical data is indeed compatible with an algebraic growth but the exponent is far smaller. The reason for this discrepancy is the assumption that all sites are equivalent which is used when approximating in the denominator of the operator-valued rate $\Gamma_k$: $\sum_{k\neq m} \frac{n_m}{|\hat{\textbf{r}}_k-\hat{\textbf{r}}_m|^\alpha}\approx F_\alpha p(\tau)$ during the derivation of Eq.\ (\ref{eq:mean-field}). This clearly does not take into account that the relaxation is hierarchical. We can improve upon this by approximating $\sum_{k\neq m} \frac{n_m}{|\hat{\textbf{r}}_k-\hat{\textbf{r}}_m|^\alpha}\approx z/\bar{l}^\alpha$ where $\bar{l}$ is the mean distance of excited atoms and $z$ the coordination number of the lattice. Clearly, $\bar{l}\propto 1/p(\tau)^{1/d}$ where $d$ is the dimension of the system. Augmenting Eq.\ (\ref{eq:mean-field}) under this assumption one obtains in the region of intermediate timescales the differential equation: $\partial_\tau p \propto p^{-2\alpha/d} R^{-2\alpha}$. This leads to an algebraically growing density of Rydberg atoms:
\begin{eqnarray}
  p(\tau)\propto (R^{-2\alpha} \tau)^\frac{d}{2\alpha+d}.\label{eq:algebraic_growth}
\end{eqnarray}
For $d=1$ and $\alpha=6$ we obtain an exponent $1/13$ which approximates well the numerical data for large $R$.

\begin{figure}
\includegraphics[width=\columnwidth]{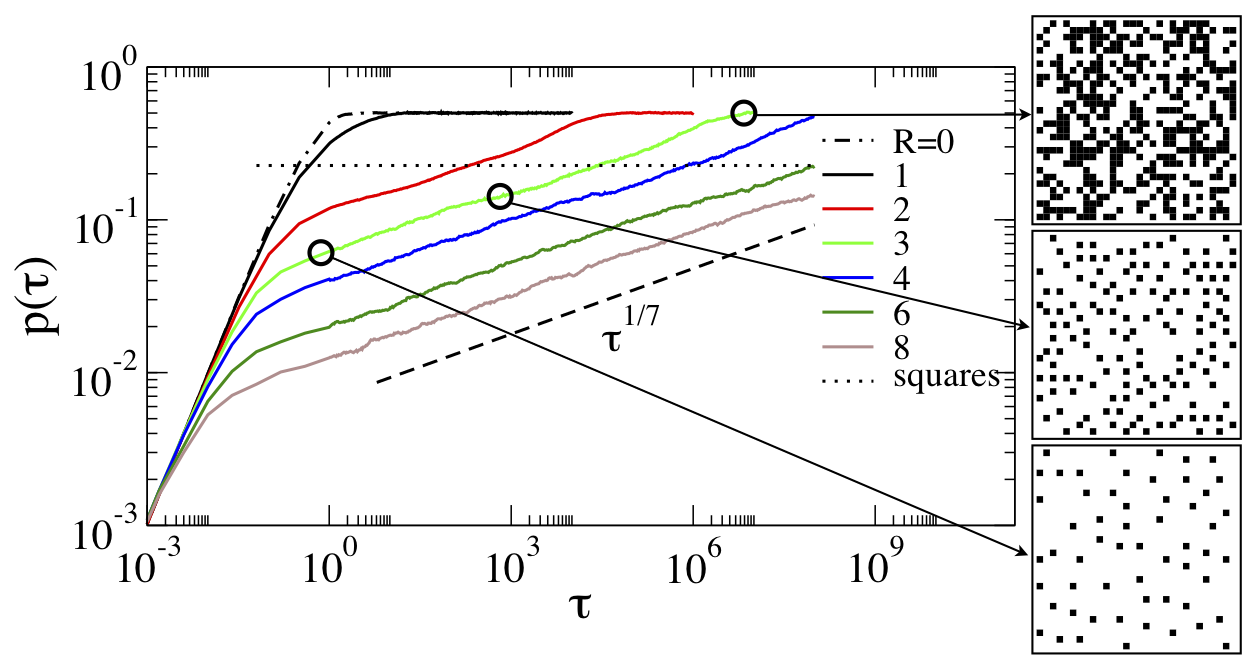}
\caption{Relaxation of the density of Rydberg atoms $p(\tau)= \sum_k p_k(\tau)/N$ in a rectangular two-dimensional system with $N=100 \times 100$, $\alpha=6$ and various values of the interaction parameter $R$. Right panels: Snapshots of the configuration of a $N=30\times30$ system at $R=3$, for the times indicated by the circles in the left part, displaying the hierarchical nature of the relaxation process.} \label{fig:2d_density}
\end{figure}

Fig. \ref{fig:2d_density} shows the corresponding simulations for two dimensions. The long and short time behavior of $p_k(\tau)$ are essentially identical to the one-dimensional case. Unlike in one dimension, however, there is no clear plateau previous to the final relaxation step. Following the earlier reasoning one could expect such plateau at a density corresponding to the maximum entropy state of the hard squares model at unity fugacity. Indeed for $R=2$ we observe a kink in that region but the feature does not prevail for larger $R$. The reason is that in two dimensions the nearest-neighbor and next-nearest-neighbor interaction energies are not as separated as in the one-dimensional case and hence the separation of time scales is not as pronounced. This prevents the formation of a clear step in the density curve.

Relaxation is also hierarchical in two dimensions as evidenced by the snapshots of configurations taken for $R=3$ in Fig.\ \ref{fig:2d_density}. This means that each excited atom is surrounded by an "excluded volume" which decreases as time passes. At intermediate times this should lead to an algebraic increase of the Rydberg density with an exponent that evaluates---according to Eq.\ (\ref{eq:algebraic_growth})---to $1/7$. Indeed this behavior is observed in the data for larger $R$. The agreement is better than in one dimension, as expected for such (augmented) mean field result.

\begin{figure}
\includegraphics[width=\columnwidth]{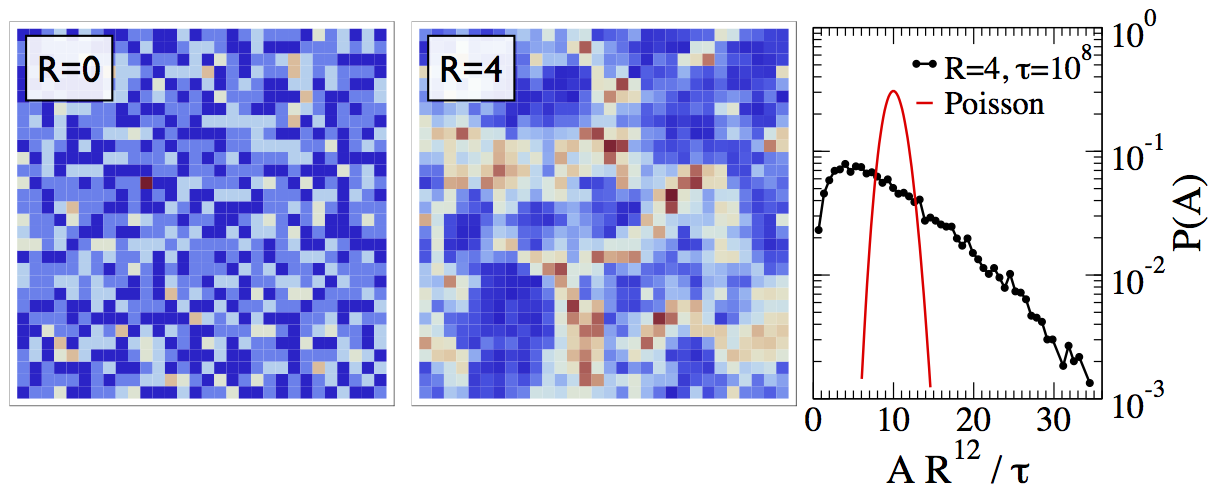}
\caption{Local activity $A_{k}$ (number of spin-flips in each site up to $\tau$) for a non-interacting system, $R=0$, after time $\tau=1$, and for an interacting one $R=4$, after time $\tau=10^8$, both of size $N=30\times30$ (red/blue indicate large/small $A_{k}$). For $R=0$ relaxation is uncorrelated in space, for $R=4$ one observes a clear spatial correlation in the dynamics, indicating dynamics heterogeneity.  The distribution of the value of the local activity in the case of $R=4,\tau=10^8$ displays fat tails (as compared to a Poisson distribution with the same average) indicative of strong dynamical fluctuations.} \label{fig:2d_activity}
\end{figure}

Fig. \ref{fig:2d_activity} illustrates the spatially heterogeneous nature of the relaxation towards equilibrium.  It shows the local activity $A_k$, defined as the total number of spin flips up to a given time $\tau$, for all sites $k$ of the lattice, for  $R=0$ and $R=4$: while in the unconstrained case the relaxation is clearly uncorrelated, for the case of $R=4$ it shows spatial segregation of fast and slow dynamics, i.e.\ dynamic heterogeneity, as in glassy systems \cite{Garrahan2002,Ediger2000}, with a broad distribution of local relaxation rates as shown by the probability of local activity, $P(A)$.

\begin{figure}
\includegraphics[width=\columnwidth]{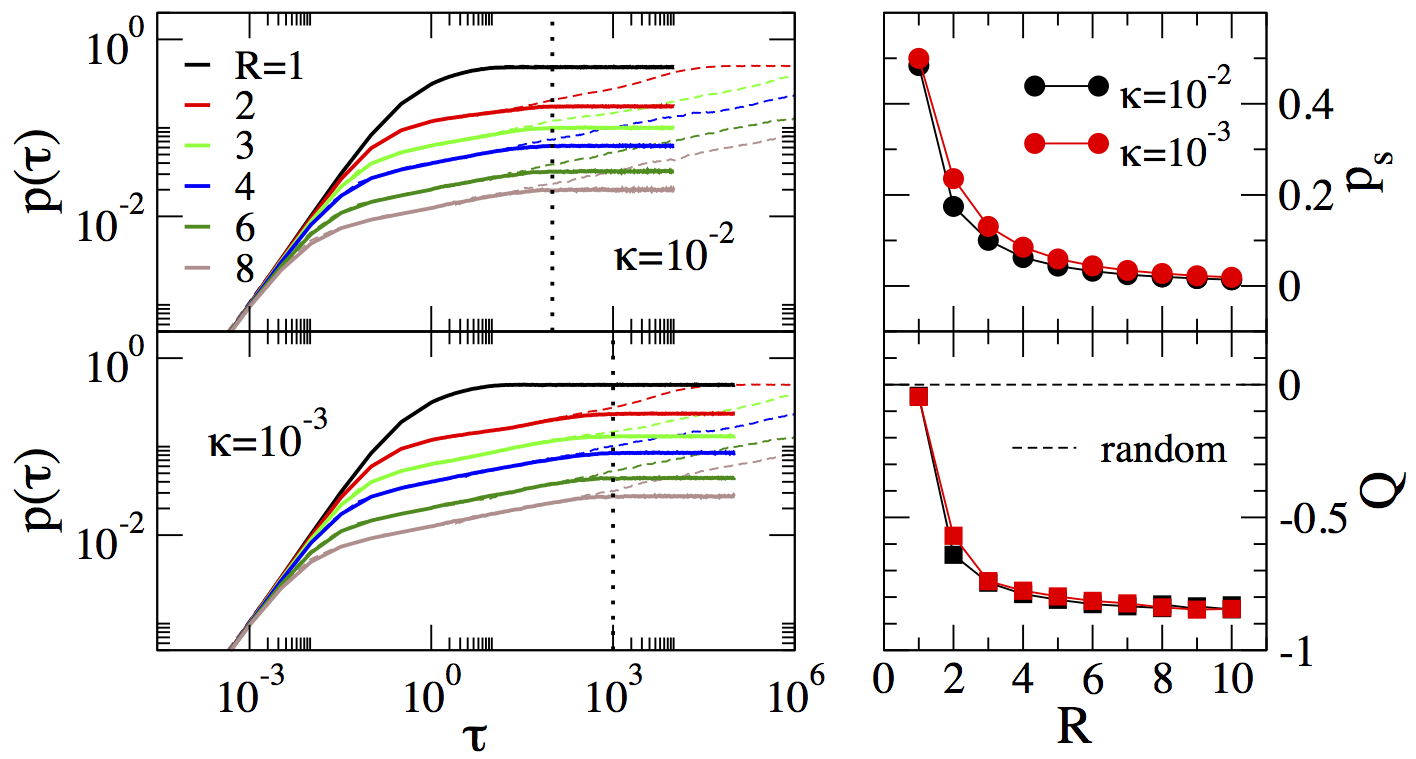}
\caption{Relaxation in a two-dimensional system of $N=100\times 100$ atoms and $\alpha=6$ with spontaneous decay $\kappa=10^{-2}$ (top left) and $10^{-3}$ (bottom left).  Dashed lines show the result for $\kappa=0$ for comparison. Top right: stationary density $p_{\rm s}$ as a function of $R$.  Bottom right: $Q$ parameter indicating correlations in the stationary state. The dashed line refers to $Q=0$ characterizing an uncorrelated and random excitation of Rydberg atoms.} \label{fig:2d_kappa}
\end{figure}

Finally, we consider the case where Rydberg atoms also decay spontaneously with a rate $\kappa$---a dissipative process which inevitably becomes relevant for sufficiently long evolution times. Assuming this decay to lead directly to the ground state \cite{Lee11} the master equation (\ref{eq:L_eff}) becomes $\partial_\tau\mu=\mathcal{L}_\mathrm{eff}\mu + \kappa \left(\sigma_-^k\mu\sigma_-^k- \frac{1}{2} \{ n_k ,\mu \} \right)$. The rate $\kappa$ breaks detailed balance, and the system in this case is driven to a non-equilibrium stationary state.  This is shown in Fig.\ \ref{fig:2d_kappa}. The initial glassy relaxation of the density in the $\kappa=0$ case above is interrupted at times $\tau \sim \kappa^{-1}$ and a non-equilibrium stationary state is reached at density $p_{\rm s}$.  For a fixed $\kappa$, $p_{\rm s}$ decreases with $R$.  More interestingly, this stationary state features pronounced spatial correlations corresponding to the excluded volume arrangements through which the glassy relaxation proceeds. This is seen by comparing the susceptibility of the number of excitations, $\chi \equiv \langle n^2 \rangle - \langle n \rangle^2$ with $n=\sum_k n_k$, to that of a random lattice gas of density $p_{\rm s}$, i.e.\, $\chi_{\rm rnd} = p_{\rm s} (1 - p_{\rm s})$.  Fig. \ref{fig:2d_kappa} shows that the corresponding $Q$-parameter,
$Q \equiv \chi/\chi_{\rm rnd} - 1$, which is often used to characterize correlations in Rydberg gases \cite{Ates06,Liebisch2005,Viteau11,Hofmann2013},
becomes progressively more negative with increasing $R$. This is a clear indication of increasing spatial correlations in the stationary state.

We have shown that the relaxation of a Rydberg lattice gas is hierarchical and exhibits glassy features such as dynamical heterogeneity due to kinetic constraints. An interesting but challenging question is to what extent these features persist beyond the limit of strong dephasing. For instance, it is not clear whether a gradual increase of the strength of the coherent driving $\Omega$ would necessarily speed up relaxation \cite{Olmos12}.  We also note that interaction Hamiltonians other than $H_{0}$ would give rise to kinetic constraints distinct from (\ref{eq:rate}) which, as in the case of anisotropic constraints \cite{Ritort2003,Olmos12}, can lead to even richer collective dynamics than the one studied here.

\acknowledgments
We acknowledge discussions with D. Petrosyan and W. Lechner. This work was supported by EPSRC Grant no.\ EP/I017828/1 and Leverhulme Trust grant no.\ F/00114/BG.

\appendix

\widetext
\section{Supplemental material: Derivation of effective master equation for $\mu$}
The dynamics of the atoms in the lattice is described by the master equation $\partial_t\rho=\mathcal{L}_0\rho+\mathcal{L}_1\rho$ with
\begin{eqnarray*}
    \mathcal{L}_0\rho &=& -i\left[H_0,\rho\right] +\gamma\sum^N_k( n_k\rho n_k-\frac{1}{2}\left\{n_k,\rho\right\})\\
    \mathcal{L}_1\rho &=& -i\Omega\sum^N_k\left[ \sigma_x^k,\rho\right].
\end{eqnarray*}
Here $n_k=(1-\sigma_z^k)/2$ and $H_0=\sum^N_k\Delta n_k+\sum_{km}\frac{V_{km}}{2}n_k n_m$ with $\sigma^k_{x,y,z}$ being the Pauli matrices. The dynamics under $\mathcal{L}_0$ is fast with respect to the slow dynamics governed by $\mathcal{L}_1$. Our aim is to derive an effective equation of motion of the slow dynamics.

Let us first look at the evolution under the fast dynamics. The operator $\mathcal{L}_0$ consists of a Hamiltonian and a dissipative part whose individual terms mutually commute. Therefore, using the abbreviation  $\mathcal{L}^k_{0,d}=\gamma( n_k\rho n_k-\frac{1}{2}\left\{n_k,\rho\right\})$ we can write the evolution of a general density matrix $\rho$ as:
\begin{eqnarray}
  e^{\mathcal{L}_0\,t} \rho=e^{-i H_0 t} \left[\bigotimes_{m} e^{\mathcal{L}^m_{0,d} t} \rho \right]e^{i H_0 t}=e^{-i H_0 t} \left[\bigotimes_{m\neq k} e^{\mathcal{L}^m_{0,d} t} \left(
                                      \begin{array}{cc}
                                        \rho^{(k)}_{\uparrow\uparrow} & \rho^{(k)}_{\uparrow\downarrow}e^{-\gamma\, t/2} \\
                                        \rho^{(k)}_{\downarrow\uparrow}e^{-\gamma\, t/2} & \rho^{(k)}_{\downarrow\downarrow} \\
                                      \end{array}
                                    \right) \right] e^{i H_0 t}.
\end{eqnarray}
In the last step we made the dissipative evolution of the $k$-th site explicit using the basis states $\left|\uparrow\right>_k$ and $\left|\downarrow\right>_k$. The symbols $\rho^{(k)}_{ij}$ represent the $N-1\times N-1$-dimensional matrices $\rho^{(k)}_{ij}=\left<i\mid\rho\mid j\right>$. We see that the off-diagonal entries of the density matrix that relate to coherences of the $k$-th spin decay exponentially.

In the long time limit the evolution under $\mathcal{L}_0$ becomes a projector on the diagonal of the density matrix $\rho$ in the product basis formed by the single particle states $\left|\uparrow\right>$, $\left|\downarrow\right>$:
\begin{eqnarray}
  \mathcal{P} \rho = \lim_{t\to\infty} e^{\mathcal{L}_0\,t} \rho= \mathrm{diag}(\rho).
\end{eqnarray}
Hence all coherences are removed and the density matrix becomes a classical state. Note that this is also not changed by the action of the coherent dynamics governed by $H_0$ as the Hamiltonian is diagonal in the product basis and coherent and dissipative evolution commute.

With the help of the projector $\mathcal{P}$ and its complement $\mathcal{Q}=1-\mathcal{P}$ we can formulate the effective evolution equation for the projected density matrix $\mu=\mathcal{P}\rho$ describing the slow evolution. This is given, to second order in $\mathcal{L}_1$, as
\begin{eqnarray}
  \partial_t \mu=\mathcal{P}\mathcal{L}_1\mu+\int_0^\infty dt\,\mathcal{P} \mathcal{L}_1 \mathcal{Q} e^{\mathcal{L}_0\,t} \mathcal{Q} \mathcal{L}_1 \mu\equiv\mathcal{L}_\mathrm{eff}\mu.
\end{eqnarray}
For the case at hand $\mathcal{P}\mathcal{L}_1\mu=0$ and $\mathcal{Q} \mathcal{L}_1 \mathcal{P}=\mathcal{L}_1 \mathcal{P}$ which leads to the expression given just before Eq.\ (1) of the main text.

Let us now evaluate the terms of $\mathcal{L}_\mathrm{eff}$. We start with
\begin{eqnarray}
  \mathcal{P}\mathcal{L}_1 e^{\mathcal{L}_0\,t} \mathcal{L}_1 \mu&=&(-i\Omega)\mathcal{P}\sum_{km}\left[\sigma_x^k,e^{\mathcal{L}_0t}[\sigma_x^m,\mu]\right]\\
  &=&-\Omega^2\mathcal{P}\sum_{km}\left[\sigma_x^k e^{\mathcal{L}_0t} \sigma_x^m \mu-\sigma_x^k e^{\mathcal{L}_0t} \mu \sigma_x^m - (e^{\mathcal{L}_0t} \sigma_x^m \mu) \sigma_x^k + (e^{\mathcal{L}_0t} \mu \sigma_x^m)\sigma_x^k\right]\\
  &=&-\Omega^2\mathcal{P}\sum_{k}\left[\sigma_x^k e^{\mathcal{L}_0t} \sigma_x^k \mu-\sigma_x^k e^{\mathcal{L}_0t} \mu \sigma_x^k - (e^{\mathcal{L}_0t} \sigma_x^k \mu) \sigma_x^k + (e^{\mathcal{L}_0t} \mu \sigma_x^k)\sigma_x^k\right]\label{eq:intermediate_result2}.
\end{eqnarray}
The simplification in the final step can be made since terms such as
\begin{eqnarray}
  \sigma_x^k e^{\mathcal{L}_0t} \sigma_x^m \mu
\end{eqnarray}
for $k\neq m$ have components with strictly off-diagonal entries when acting on diagonal density matrices, e.g.
\begin{eqnarray}
\sigma_x^k \sigma_x^m \rho=\sigma_x^k\otimes \sigma_x^m\left(
                             \begin{array}{cccc}
                               \rho^{(k,m)}_{\uparrow\uparrow\uparrow\uparrow} & 0 & 0 & 0 \\
                               0 & \rho^{(k,m)}_{\uparrow\uparrow\downarrow\downarrow} & 0 & 0 \\
                               0 & 0 & \rho^{(k,m)}_{\downarrow\downarrow\uparrow\uparrow} & 0 \\
                               0 & 0 & 0 & \rho^{(k,m)}_{\downarrow\downarrow\downarrow\downarrow} \\
                             \end{array}
                           \right)=\left(
                             \begin{array}{cccc}
                               0 & 0 & 0 & \rho^{(k,m)}_{\downarrow\downarrow\downarrow\downarrow} \\
                               0 & 0 & \rho^{(k,m)}_{\downarrow\downarrow\uparrow\uparrow} & 0 \\
                               0 & \rho^{(k,m)}_{\uparrow\uparrow\downarrow\downarrow} & 0 & 0 \\
                                \rho^{(k,m)}_{\uparrow\uparrow\uparrow\uparrow}& 0 & 0 &0 \\
                             \end{array}
                           \right),\nonumber
\end{eqnarray}
which therefore equate to zero under the action of the projector $\mathcal{P}$.

Let us now calculate the explicit form of the first term of Eq.\ (\ref{eq:intermediate_result2}):
\begin{eqnarray}
  \sigma_x^k \mu &=& \sigma_x^k\left(
     \begin{array}{cc}
       \rho^{(k)}_{\uparrow\uparrow} & 0 \\
       0 & \rho^{(k)}_{\downarrow\downarrow} \\
     \end{array}
   \right)=\left(
     \begin{array}{cc}
       0 & \rho^{(k)}_{\downarrow\downarrow} \\
       \rho^{(k)}_{\uparrow\uparrow} & 0 \\
     \end{array}
   \right)\\
     e^{\mathcal{L}_0t} \sigma_x^k \mu &=& e^{-i H_0 t}\left(
     \begin{array}{cc}
       0 & \rho^{(k)}_{\downarrow\downarrow} e^{-\gamma t/2} \\
       \rho^{(k)}_{\uparrow\uparrow} e^{-\gamma t/2} & 0 \\
     \end{array}
   \right)e^{i H_0 t}\\
   &=&\left(
     \begin{array}{cc}
       0 & \rho^{(k)}_{\downarrow\downarrow} e^{-\gamma t/2} e^{-it (\Delta+\sum_m V_{km} n_m)} \\
       \rho^{(k)}_{\uparrow\uparrow} e^{-\gamma t/2} e^{it (\Delta+\sum_m V_{km} n_m)}& 0 \\
     \end{array}
   \right)\\
   \sigma_x^k e^{\mathcal{L}_0t} \sigma_x^k \mu &=& \left(
     \begin{array}{cc}
        \rho^{(k)}_{\uparrow\uparrow} e^{-\gamma t/2} e^{it (\Delta+\sum_m V_{km} n_m)}&0 \\
       0 & \rho^{(k)}_{\downarrow\downarrow} e^{-\gamma t/2} e^{-it (\Delta+\sum_m V_{km} n_m)} \\
     \end{array}
   \right)\\
   &=& e^{-\gamma t/2} e^{it \sigma^k_z(\Delta+\sum_m V_{km} n_m)} \mu.
\end{eqnarray}
Furthermore we find
\begin{eqnarray}
e^{\mathcal{L}_0t} (\mu\sigma_x^k) \sigma_x^k &=&  e^{-\gamma t/2} e^{-it \sigma^k_z(\Delta+\sum_m V_{km} n_m)} \mu\\
\sigma^k_x e^{\mathcal{L}_0t} (\mu \sigma_x^k) &=& \sigma^k_x e^{-\gamma t/2} e^{-it \sigma^k_z(\Delta+\sum_m V_{km} n_m)} \mu\,\sigma^k_x\\
e^{\mathcal{L}_0t} (\sigma_x^k \mu) \sigma_x^k &=& \sigma_x^k e^{-\gamma t/2} e^{it\sigma^k_z (\Delta+\sum_m V_{km} n_m)} \mu\,\sigma^k_x.
\end{eqnarray}
With this result and abbreviating $Y_k(t)=2 e^{-\gamma t/2} \cos(\sigma^k_z[\Delta+\sum_m V_{km} n_m]t)$ we find
\begin{eqnarray}
  \mathcal{L}_\mathrm{eff}\mu=-\Omega^2 \int_0^\infty dt\, \sum_k \left[Y_k(t)\mu -\sigma_x^k Y_k(t) \mu \sigma_x^k\right],
\end{eqnarray}
and using
\begin{eqnarray}
  \int_0^\infty dt\, Y_k(t)= 4\frac{\gamma}{\gamma^2+\left(2\Delta+2\sum_m V_{km} n_m\right)^2}
\end{eqnarray}
we find
\begin{eqnarray}
  \mathcal{L}_\mathrm{eff}\mu=\sum_k4\frac{\gamma\Omega^2}{\gamma^2+\left(2\Delta+2\sum_m V_{km} n_m\right)^2}\left[\sigma_x^k \mu \sigma_x^k-\mu\right]=\frac{4\Omega^2}{\gamma} \sum_k\Gamma_k \left[\sigma_x^k \mu \sigma_x^k-\mu\right]
\end{eqnarray}
which corresponds to Eq.\ (1) in the manuscript.

\end{document}